\documentclass[12pt]{article}

\begin{document}
\begin{center}
{\it NOETHER SYMMETRIES FOR CHARGED PARTICLE MOTION UNDER A MAGNETIC MONOPOLE AND GENERAL ELECTRIC FIELDS}

\vskip .7cm

{\bf F. HAAS$^1$}

\vskip .3cm

Departamento de Matem\'atica Aplicada e Computacional \\ Laborat\'orio Nacional de Computa\c{c}\~ao
Cient\'{\i}fica \\ Av. Get\'ulio Vargas, 333\\
Petr\'opolis, 25651-070, Brazil \\
\strut$^1$ E-mail: ferhaas@lncc.br
\end{center}

\vskip 1cm

\date{\strut}

\noindent{{\bf ABSTRACT:} {\it The search for Noether point symmetries for
non-re\-la\-ti\-vis\-tic charged particle motion is reduced to the solution
for a set of two coupled, linear partial differential equations for
the electromagnetic field. These equations are completely solved
when the magnetic field is produced by a fixed magnetic monopole.
The result is applied to central electric field cases, in
particular to the time-dependent monopole-oscillator problem. As
an additional example of the theory, we found all Noether
point symmetries and invariants for a constant magnetic field  and
a time-dependent harmonic electric field with a forcing term.}}


\vskip .7cm

\section{Introduction}

The charge-monopole problem is a classical subject in Physics. In
the present work, in particular, we consider symmetries and
conservation laws for the Lorentz equations of the form
\begin{equation}
\label{eqLorentz}
\ddot{\bf r} = -
\nabla\,V({\bf r},t) + g\frac{\dot{\bf r}\times{\bf r}}{r^3} \,,
\end{equation}
applying Noether's theorem. Here, ${\bf r} = (x,y,z)$ is the
position vector in $R^3$. The system (\ref{eqLorentz}) describes
three-dimensional, non-relativistic charged particle motion
under an electric field ${\bf E} = - \nabla\,V({\bf r},t)$ and a
fixed magnetic monopole field with strength $g$. However, the
scope of our work is more general, since our formalism apply to
general electromagnetic fields. We make an effort
to extend the results of \cite{H1, H2}, where all the Noether and
Lie points symmetries for two-dimensional, non-relativistic
charged particle motions were found. Unlike the planar case, the
fully three-dimensional case seems to be not accessible to a
complete solution. Hence, we focus mainly on the magnetic monopole
field case, which is amenable to complete calculations.

Let us review the state of the art on the search for constants of
motion for the charge-monopole system. In the simpler case when
the electric force is central and time-independent ($V = V(r)$, $r
= (x^2 + y^2 + z^2)^{1/2}$), the Lorentz equations admit the
vector first integral
\begin{equation}
\label{re}
{\bf D} = {\bf r}\times\dot{\bf r} - g\hat{r} \,,
\end{equation}
where $\hat{r}$ is the unit vector in the radial direction.  The
vector ${\bf D}$, the so-called Poincar\'e vector, was used by
Poincar\'e \cite{re} to obtain the exact solution for the motion
when only the magnetic monopole is present ($V = 0$). It should be
mentioned that the Poincar\'e vector survives as a constant of
motion even if there is an explicit time-dependence of the scalar
potential.

More recently \cite{McIntosh}, it were considered the scalar potentials
\begin{equation}
\label{oscilador}
V = \frac{\omega_{0}^{2}{r^2}}{2} +
\frac{g^2}{2r^2}
\end{equation}
and
\begin{equation}
\label{Kepler}
V = -\frac{\mu_0}{r} + \frac{g^2}{2r^2} \,,
\end{equation}
where $\omega_0$ and $\mu_0$ are numerical constants. All bounded
trajectories are periodic \cite{McIntosh} when the scalar
potential is given by (\ref{oscilador}) or (\ref{Kepler}). For the
potential (\ref{oscilador}), linked to the isotropic harmonic
oscillator, there is a tensor conserved quantity, whose components, using complex notation, are
\begin{equation}
\label{tensor}
T_{ij} = (\dot{u_i} + i\omega_{0}u_i)(\dot{u_j} -
i\omega_{0}u_j) \,,
\end{equation}
where ${\bf u} = {\bf D}\times\hat{r}$. In the case of the potential (\ref{Kepler}), related to the Coulomb or Kepler
forces, we have the vector constant of motion
\begin{equation}
\label{Runge}
{\bf F} = {\bf D}\times\dot{\bf r} +
\mu_{0}\hat{r} \,,
\end{equation}
a generalization of the Laplace-Runge-Lenz vector for the Kepler
problem. Both (\ref{tensor}) and (\ref{Runge}) are constants of
motion in the form of quadratic functions of the velocity. It can
be demonstrated  \cite{Thompson} that (\ref{oscilador}) and
(\ref{Kepler}) are among the few time-independent central potentials for which (\ref{eqLorentz}) has quadratic
integrals other than the energy. At the quantum level, there is degeneracy of the spectra for these potentials, in
connection \cite{Labelle} with the invariance
algebra $su(2)\oplus\,su(2)$.

In contradistinction to these earlier works, here we consider the
effects of the superposition of a non-central, time-dependent
electric force on the motion of charged particles under a fixed
magnetic monopole field. An immediate result of the presence
of non-central electric fields is the non-conservation of the
Poincar\'e vector (\ref{re}). However, at least for particular
forms of $V({\bf r},t)$, it can be expected that some conservation
law is present. To address the question, we pursue here the
analysis of Noether point symmetries.

As an additional example of the theory, we found all Noether point
symmetries and invariants for a constant magnetic field and a
time-dependent harmonic electric field with a forcing term. In
this case there is a 12-parameter or a 8-parameter group of
symmetries, according to a condition on the parameters specifying
the electromagnetic field. Apart from it's obvious physical
significance, the extra example shows how to apply the general
theory of this work to situations for which the electromagnetic field has a more particular form, known in advance.

The article is structured as follows. In Section II, we
investigate the Noether point symmetries for non-relativistic
charged particle motion under general electromagnetic fields. The
whole problem is reduced to a set of two coupled, linear partial
differential equations for the electromagnetic field. In Section
III, we completely solve this system of equations when the
magnetic field is in the form of a fixed magnetic monopole field.
Two classes of electric field are determined, related to quadratic
or linear constants of motion. These electric fields can be freely
superimposed to the magnetic monopole field, with no harm on the
existence of Noether point symmetries. In Section IV, the previous
results are applied to the central electric field cases. In
particular, we study the time-dependent monopole-oscillator
system. In Section V, the case of a constant magnetic field is
analyzed. Unlike the magnetic monopole problem, here we will not
try to find the general class of admissible electric fields.
Instead, we focus only on linear electric fields, which are
amenable to exact calculations. Section VI is dedicated to the
conclusion.

\section{Noether point symmetries}

The necessary and sufficient condition \cite{Sarlet} for a vector
field
\begin{equation}
\label{NoetherG} G = \tau({\bf r},t)\frac{\partial}{\partial t} +
{\bf n }({\bf r},t)\cdot\frac{\partial}{\partial{\bf r}}
\end{equation}
to be a generator of Noether point symmetries for the action functional
\begin{equation}
S[{\bf r}(t)] = \int^{t_1}_{t_0} L({\bf r},\dot{\bf r},t)\,dt
\,,
\end{equation}
where $L({\bf r},\dot{\bf r},t)$ is the Lagrange function,
is the existence of a function $F({\bf r},t)$ such that
\begin{equation}
\label{NoetherC}
\tau\frac{\partial L}{\partial t} + {\bf n }\cdot\frac{\partial
L}{\partial{\bf r}} + (\dot{\bf n } - \dot\tau\dot{\bf r})\cdot\frac{\partial
L}{\partial\dot{\bf r}} + \dot\tau\,L = \dot F({\bf r},t) \,.
\end{equation}
Notice that the generator $G$ in (\ref{NoetherG}) does not include
derivatives of the coordinates, so that dynamical symmetries are
not being considered here.

Associated to the symmetries satisfying (\ref{NoetherC}), there is
a first integral of the form
\begin{equation}
\label{NoetherI} I = (\frac{\partial
L}{\partial\dot{\bf r}}\cdot\dot{\bf r} - L)\,\tau - {\bf n }\cdot\frac{\partial
L}{\partial\dot{\bf r}} + F \,,
\end{equation}
conserved along the trajectories of the Euler-Lagrange equations.

Noether's theorem is applicable to any Lagrangian system, as is
the case for non-relativistic motion of a charged particle under a
general electromagnetic field. Introduce vector ${\bf A}({\bf
r},t)$ and scalar $V({\bf r},t)$ potentials, so that the
Lagrangian is given by
\begin{equation}
L = \frac{1}{2}{\dot{\bf r}}^2 + {\bf A}({\bf r},t)\cdot\dot{\bf r} - V({\bf r},t)
\,.
\end{equation}
The corresponding electromagnetic fields are
\begin{equation}
\label{EMF}
{\bf B} = \nabla\times{\bf A} \quad , \quad {\bf E}
= - \nabla\,V - \partial{\bf A}/\partial t \quad .
\end{equation}
Inserting $L$ in the symmetry condition (\ref{NoetherC}), we find
a polynomial form of the velocity components. The coefficient of
each monomial should be zero. Such a prescription results in a
system of linear partial differential equations determining both
the symmetry generator and the vector and scalar potentials.
Actually, we will show in the continuation that it is possible to
reduce the discussion to the electromagnetic field only, without
any mention to the electromagnetic potentials.

Putting forward the calculation of Noether symmetries, the
monomial of  third order on the velocity gives, using component
notation,
\begin{equation}
\label{pri}
\frac{\partial\tau}{\partial r_i} = 0 \,,
\end{equation}
thus implying
\begin{equation}
\label{x1}
\tau = {\rho^{2}}(t) \,,
\end{equation}
where $\rho(t)$ is an arbitrary function of time. Now, the
monomial of  second order on the velocity imposes
\begin{equation}
\frac{\partial n_i}{\partial
r_j} + \frac{\partial n_j}{\partial r_i} - 2\,\delta_{ij}\rho\dot\rho = 0 \,,
\end{equation}
where the Kronecker delta was used. The solution is
\begin{equation}
\label{x2} {\bf n} = \rho\dot\rho{\bf r} +
{\bf\Omega}(t)\times{\bf r} + {\bf a}(t) \,,
\end{equation}
where ${\bf\Omega}(t)$ and ${\bf a}(t)$ are arbitrary vector functions of time.

Assembling (\ref{x1}-\ref{x2}), we conclude that the most general
form of the Noether point symmetry generator is
\begin{equation}
\label{geral}
G =
\rho^{2}(t)\frac{\partial}{\partial t} + (\rho\dot\rho{\bf r} +
{\bf\Omega}(t)\times{\bf r} + {\bf a}(t))\cdot\frac{\partial}{\partial{\bf r}}
\,,
\end{equation}
for arbitrary $\rho(t)$, ${\bf\Omega}(t)$ and ${\bf a}(t)$. The
resulting symmetries include a generalized rescaling, a
time-dependent rotation and a time-dependent space translation. Up
to this point, there is no restriction on the electromagnetic
field. Notice that (\ref{geral}) is the proper extension of the
Noether point symmetries generator for two-dimensional
non-relativistic charged particle motion derived in \cite{H1}.

The remaining equations implied by the symmetry condition (\ref{NoetherC}) are
\begin{eqnarray}
\label{F1}
\nabla F = G{\bf A} + \rho\dot\rho{\bf A} + {\bf
A}\times{\bf\Omega} + \partial{\bf n}/\partial t \,,\\
\label{F2}
\partial
F/\partial t = - G\,V - 2\rho\dot\rho V + {\bf A}\cdot\partial{\bf n}/\partial
t \,,
\end{eqnarray}
in which the form (\ref{geral}) was taken into account. Also, we
have used the definition
\begin{equation}
G\,W =
\rho^{2}(t)\frac{\partial\,W}{\partial t} + (\rho\dot\rho{\bf r} +
{\bf\Omega}(t)\times{\bf r} + {\bf a}(t))\cdot\frac{\partial\,W}{\partial{\bf r}}
\,,
\end{equation}
valid for a generic function $W = W({\bf r},t)$.

Equations (\ref{F1}--\ref{F2}) have a solution $F$ if and only if
\begin{equation}
\label{x3}
\frac{\partial^{2}F}{\partial r_{i}\partial r_{j}} =
\frac{\partial^{2}F}{\partial r_{j}\partial r_{i}} \quad , \quad
\frac{\partial^{2}F}{\partial r_{i}\partial t} = \frac{\partial^{2}F}{\partial
t\partial r_{i}} \,.
\end{equation}
Using (\ref{F1}--\ref{F2}) in (\ref{x3}), we obtain
\begin{eqnarray}
\label{basicB}
G{\bf B} &=& - 2\rho\dot\rho{\bf B} - 2\dot{\bf\Omega} +
{\bf\Omega}\times{\bf B} \,,\\
\label{basicE}
G{\bf E} &=& - 3\rho\dot\rho{\bf E}
+ {\bf\Omega}\times{\bf E} + {\bf B}\times\frac{\partial{\bf n}}{\partial t} +
\frac{\partial^{2}{\bf n}}{\partial t^2} \,,
\end{eqnarray}
involving only the electromagnetic fields, not the electromagnetic
potentials. Hence, the choice of gauge does not have any influence
in the search for Noether point symmetries.

Equations (\ref{basicB}--\ref{basicE}) are the fundamental
equations for the determination of Noether point symmetries for
non-relativistic charged particle motion. It is a system of
coupled, linear partial differential equations for the fields
${\bf E}$ and ${\bf B}$, involving the functions $\rho$,
${\bf\Omega}$ and ${\bf a}$ that define the generator $G$. The
system to be satisfied by the electromagnetic fields is the proper
three-dimensional extension of the system (34-36) found in
\cite{H1} in the planar charged particle motion case. As long as
we know, this is the first time equations
(\ref{basicB}-\ref{basicE}) are explicitly written.

Unlike the two-dimensional case, it seems that the complete
solution of (\ref{basicB}-\ref{basicE}) is not available. Thus, a
complete classification of the electromagnetic fields for which
the three-dimensional non-relativistic charged particle motion
possesses Noether point symmetries is not known. The technical
difficulties arising in the fully three-dimensional case are
related to the problem of finding the canonical group coordinates
associated to the generator (\ref{geral}), for arbitrary functions
$\rho(t)$, ${\bf\Omega}(t)$ and ${\bf a}(t)$. Nevertheless,
equations (\ref{basicB}-\ref{basicE}) are useful in the
investigation of specific electromagnetic fields. For instance, in
what follows we show that when the magnetic field is produced by a
fixed magnetic monopole it is possible to obtain the general
solution for (\ref{basicB}-\ref{basicE}).

Of course, the electromagnetic fields satisfying
(\ref{basicB}-\ref{basicE}) must also comply with Maxwell
equations, a condition that is not immediately assured. The
non-ho\-mo\-ge\-neous Maxwell equations may always be satisfied by a
convenient choice of charge and current densities. There remains
the Gauss law for magnetism and Faraday's law. Gauss law is
removed when magnetic monopoles are present. In these cases, Faraday's law is the only extra requirement.

Once the system (\ref{basicB}--\ref{basicE}) is solved for the
electromagnetic field, the Noether invariant follows from
(\ref{NoetherI}), reading
\begin{equation}
\label{basicI}
I = (\dot{\bf r}^{2}/2 + V)\rho^2 - (\dot{\bf r}
+ {\bf A})\cdot{\bf n} + F \,.
\end{equation}
The invariant $I$ is a quadratic polynomial on the velocities when
$\rho\neq 0$. Otherwise, for $\rho = 0$, $I$ is a linear
polynomial on the velocities. It is apparent from (\ref{basicI})
that the Noether invariant needs the electromagnetic potentials as
well as the function  $F({\bf r},t)$ obtained from
(\ref{F1}--\ref{F2}). However, as we will see in concrete
examples, the form of $I$ is in practice independent of gauge, as
expected.

Equations (\ref{basicB}-\ref{basicE}), restricted to planar
motions,  were completely solved in \cite{H1}. In the next Section we pursue a less ambitious program, taking ${\bf
B}$ as the field of a  fixed magnetic monopole and studying the consequences of this choice on
the generator $G$ and on the electric field.

\section{Magnetic monopole}

Here we apply the formalism of Section II to the case
of a magnetic monopole fixed at origin and with strength $g$,
\begin{equation}
\label{Bmonop}
{\bf B} = \frac{g\,{\bf
r}}{r^3} \,.
\end{equation}
Inserting (\ref{Bmonop}) in (\ref{basicB}), the result is a
condition on the functions $\rho$, ${\bf \Omega}$ and ${\bf a}$
composing the symmetry,
\begin{equation}
\label{condS} g\,(r^2{\bf a} - 3{\bf
a}\cdot{\bf r}{\bf r})r^{-5} - \dot{\bf\Omega} = 0 \,.
\end{equation}
Notice that $\rho$ is not present, remaining arbitrary. Equation
(\ref{condS}) is identically satisfied if and only if
\begin{equation}
{\bf a} = \dot{\bf\Omega}
= 0 \,.
\end{equation}
Consequently, the generator of Noether point symmetries is specified by
\begin{equation}
\label{ER3SO3}
G =
\rho^{2}\frac{\partial}{\partial t} + (\rho\dot\rho{\bf r} +
{\bf\Omega}\times{\bf r})\cdot\frac{\partial}{\partial{\bf r}} \,.
\end{equation}
In other words,
\begin{equation}
G = G_\rho + {\bf\Omega}\cdot{\bf L} \,,
\end{equation}
where
\begin{equation}
\label{ER3}
G_\rho =
\rho^{2}\frac{\partial}{\partial\,t} + \rho\dot\rho{\bf
r}\cdot\frac{\partial}{\partial{\bf r}} \,,
\end{equation}
is the generator of quasi-invariance transformations \cite{GCT, Munier}, ${\bf\Omega}$ is from now on a constant vector and ${\bf L}
= (L_1,L_2,L_3)$ is defined by
\begin{equation}
L_1 = y\frac{\partial}{\partial z} - z\frac{\partial}{\partial y} \,, \quad L_2 =
z\frac{\partial}{\partial x} - x\frac{\partial}{\partial z} \,, \quad L_3 =
x\frac{\partial}{\partial y} - y\frac{\partial}{\partial x} \,.
\end{equation}
We recognize $L_1$, $L_2$ and $L_3$ as the generators of the $so(3)$ algebra.

The electric fields compatible with Noether point symmetry satisfy
(\ref{basicE}). With generator given by (\ref{ER3SO3}), this
equation for ${\bf E}$ reads
\begin{equation}
\label{Emono} G{\bf E} = - 3\rho\dot\rho{\bf E} +
{\bf\Omega}\times{\bf E} + (\rho{\buildrel\cdots\over\rho} +
3\dot\rho\ddot\rho){\bf r} \,.
\end{equation}
Notice that the strength $g$ does not appear on (\ref{Emono}).

Before considering (\ref{Emono}) in the general case, it is
instructive to examine first the case ${\bf E} = 0$, in which only
the magnetic monopole is present. For no electric field,
(\ref{Emono}) reduces to
\begin{equation}
\rho{\buildrel\cdots\over\rho} + 3\dot\rho\ddot\rho = 0 \,,
\end{equation}
whose general solution is
\begin{equation}
\label{gera1}
\rho^2 = c_1 + c_{2}t
+ c_{3}t^2 \,,
\end{equation}
where $c_1$, $c_2$ and $c_3$ are arbitrary constants. We conclude
by the existence of three symmetry generators,
\begin{equation}
\label{g1}
G_1 = \frac{\partial}{\partial t} \,,\quad G_2 =
t\frac{\partial}{\partial t} + \frac{{\bf r}}{2}\cdot\frac{\partial}{\partial{\bf r}} \,,\quad G_3 =
t^{2}\frac{\partial}{\partial t} + t{\bf r}\cdot\frac{\partial}{\partial{\bf r}} \,.
\end{equation}
These generators are associated, respectively, to time
translation, self-si\-mi\-la\-ri\-ty and conformal transformations,
composing the $so(2,1)$ algebra, with commutation relations
\begin{equation}
\label{so21}
\left[G_1,G_2\right] = G_1 \,, \quad
\left[G_1,G_3\right] = 2G_2 \,, \quad \left[G_2,G_3\right] = G_3 \,.
\end{equation}
Therefore, the problem where only the magnetic monopole is present
is endowed with the $SO(2,1)\times\,SO(3)$ group of Noether point
symmetries. Such a result was already obtained by Jackiw
\cite{Jackiw}, using dynamical Noether transformations, and by
Moreira et al. \cite{Moreira}, using Lie point symmetries and no
variational formulation. Lie's approach has the advantage of no
necessity of electromagnetic potentials, which are always singular
when magnetic monopoles are present. However, as seen in Section
II, the basic equations (\ref{basicB}--\ref{basicE}) can be
formulated in terms of the electromagnetic field only. Moreover,
we shall see in practice the  gauge invariance of the Noether
invariant. Our procedure is simpler than, for instance, the use of fiber bundles to avoid the singularity of the
vector potential \cite{Soko}.

The solution for (\ref{Emono}) comprises two categories, one for
$\rho \neq 0$ and the other for $\rho = 0$. Accordingly,
(\ref{basicI}) shows that each class of solution is associated to
quadratic or linear constants of motion, respectively.

\subsection{The case $\rho \neq 0$}

Using the method of characteristics, we find that when
$\rho\neq\,0$ the solution for (\ref{Emono}) is given by
\begin{equation}
\label{Eperm1}
{\bf E} =
\frac{\ddot\rho}{\rho}{\bf r} +
\frac{1}{\rho^4}{\bf\Omega}\times({\bf\Omega}\times{\bf r}) +
\frac{1}{\rho^3}R(\Omega\bar{t})\cdot\bar{\bf E}(\bar{\bf r}) \,,
\end{equation}
where we have used the definitions
\begin{equation} \bar{t} =
\int^{t}\,d\tau/\rho^{2}(\tau) \,,\quad \bar{\bf r} =
\frac{1}{\rho}R^{T}(\Omega\bar{t})\cdot{\bf r} \,,
\end{equation}
for $R(\Omega\bar{t})$ the rotation matrix about ${\bf\Omega}$ by
an angle $\Omega\bar{t}$. The symbol $T$ is for the transpose.
Also, in  (\ref{Eperm1}) $\bar{\bf E}$ is an arbitrary vector
function of the indicated argument. For ${\bf\Omega} =
(0,0,\Omega)$, the explicit form of the rotation matrix is
\begin{equation}
R(\Omega\bar{t}) = \pmatrix{ \cos \Omega\bar{t}&- \sin
\Omega\bar{t}&0\cr \sin \Omega\bar{t}&\cos \Omega\bar{t}&0\cr 0&0&1\cr} \,.
\end{equation}

The reader can directly verify that (\ref{Eperm1}) satisfy
(\ref{Emono}). For this check, the relations
\begin{eqnarray}
\label{bart}
\frac{\partial}{\partial\bar{t}} &=& \rho^{2}\frac{\partial}{\partial t} +
(\rho\dot\rho{\bf r} + {\bf\Omega}\times{\bf r})\cdot\frac{\partial}{\partial{\bf
r}} \,,\\ \label{barr} \frac{\partial}{\partial\bar{\bf r}} &=&
\rho\,R^{T}(\Omega\bar{t}(t))\cdot\frac{\partial}{\partial{\bf r}}
\end{eqnarray}
are useful. Moreover, (\ref{ER3SO3}) and (\ref{bart}) shows that
\begin{equation}
G = \partial/\partial\bar{t}
\end{equation}
that is, $\bar{\bf r}$ and $\bar{t}$ are canonical group
coordinates for the Noether point symmetries with $\rho \neq 0$, so that $G$ is  the generator of translations along
$\bar{t}$.

We see, on the electric field, the presence of the arbitrary
functions $\rho$ and $\bar{\bf E}$. However, for the vector field
(\ref{Eperm1}) to qualify as a true electric field, it must
complain with Faraday's law,
\begin{equation}
\nabla\times{\bf E} +
\frac{\partial{\bf B}}{\partial t} = 0 \,,
\end{equation}
which imposes, for
${\bf E}$ as in (\ref{Eperm1}),
\begin{equation}
\label{pot}
R(\Omega\bar{t})\cdot(\bar{\nabla}\times\bar{\bf E}(\bar{\bf r})) = 0 \,,
\end{equation}
where $\bar{\nabla} = \partial/\partial\bar{\bf r}$. As the
rotation matrix $R$ is non singular, the only way of satisfying
(\ref{pot}) is
\begin{equation}
\label{Epot}
\bar{\bf E} = - \bar\nabla\,U(\bar{\bf r})
\end{equation}
for some function $U(\bar{\bf r})$. The remaining Maxwell
equations (with exception of Gauss law for magnetism, which has
been ruled out) can be satisfied by an appropriate choice of
charge $n_q$ and current ${\bf J}_q$ densities,
\begin{eqnarray}
n_q &\equiv& \nabla\cdot{\bf E} = 3\frac{\ddot\rho}{\rho} -
2\frac{{\bf\Omega}^2}{\rho^4} - \frac{1}{\rho^2}\nabla^{2}U \,,\\
{\bf J}_q
&\equiv& \nabla\times{\bf B} - \partial{\bf E}/\partial\,t =
(\frac{\dot\rho\ddot\rho}{\rho^2} - \frac{\buildrel\cdots\over\rho}{\rho}){\bf r}
+ 4\frac{\dot\rho}{\rho^5}{\bf\Omega}\times({\bf\Omega}\times{\bf r}) \nonumber
\\ &\strut& \qquad\qquad\qquad\quad\quad - 2\frac{\dot\rho}{\rho^3}\nabla\,U +
\frac{1}{\rho^2}\nabla\frac{\partial\,U}{\partial t} \,.
\end{eqnarray}

To resume, taking into account (\ref{Epot}) and also (\ref{barr}), it follows that
\begin{equation}
\label{Eperm2}
{\bf E} = \frac{\ddot\rho}{\rho}{\bf r} +
\frac{1}{\rho^4}{\bf\Omega}\times({\bf\Omega}\times{\bf r}) -
\frac{1}{\rho^2}\nabla\,U(\bar{\bf r}) \,,
\end{equation}
is the general form of the admissible electric fields, compatible
with Noether point symmetries (with $\rho \neq 0$) and a magnetic
monopole field. The electric field depend on the arbitrary
functions $\rho$ and $U$, as well as on the constant vector
${\bf\Omega}$. Central fields are obtained as particular cases
taking ${\bf\Omega} = 0$ and $U = U(\bar{r})$, where $\bar{r}$ is
the norm of $\bar{\bf r}$.

The Noether invariant (\ref{basicI}) require the electromagnetic
potentials, as well as the function $F$ solution of
(\ref{F1}--\ref{F2}). In full generality, the electromagnetic
potentials are given by
\begin{eqnarray}
\label{Amono}
{\bf A} &=& \frac{g\,z}{r(x^2 + y^2)}(y,-x,0) + \nabla\lambda({\bf r},t) \,, \\
\label{Vperm}
V &=& - \frac{\ddot\rho\,r^2}{2\rho} +
\frac{1}{2\rho^4}({\bf\Omega}\times{\bf r})^2 + \frac{1}{\rho^2}U(\bar{\bf r}) -
\frac{\partial\lambda}{\partial\,t}({\bf r},t) \,,
\end{eqnarray}
where $\lambda({\bf r},t)$ is an arbitrary gauge function.
Inserting the electromagnetic potentials into (\ref{F1}-\ref{F2}),
there results a system whose solution is
\begin{equation}
F = \frac{1}{2}(\dot\rho^2 + \rho\ddot\rho)\,r^2 +
g\,r\frac{(\Omega_{1}x + \Omega_{2}y)}{(x^2 + y^2)}\,G\lambda \,,
\end{equation}
so that the Noether first integral, from (\ref{basicI}), is
\begin{equation}
\label{Iquadra}
I = \frac{1}{2}(\rho\dot{\bf r} -
\dot\rho{\bf r} - {\bf\Omega}\times{\bf r}/\rho)^2 + U(\bar{\bf r}) +
g{\bf\Omega}\cdot\hat{r} \,.
\end{equation}
As anticipated, $I$ is not dependent on the gauge function $\lambda$.

\subsection{The case $\rho = 0$}

For $\rho = 0$, (\ref{Emono}) reduces to
\begin{equation}
\label{Erho0}
{\bf\Omega}\times{\bf r}\cdot\frac{\partial{\bf E}}{\partial{\bf r}} =
{\bf\Omega}\times{\bf E}  \,.
\end{equation}
Since ${\bf \Omega}$ is a constant vector, there is no loss of
generality if we set ${\bf\Omega} = (0,0,\Omega)$ in
(\ref{Erho0}), with $\Omega \neq 0$, so that $G$ becomes $L_3$,
the generator of rotations about the $z$ axis. In this situation,
the general solution for (\ref{Erho0}) is
\begin{equation}
\label{eee}
{\bf E} =
E_{r}(r,\theta,t)\hat{r} + E_{\theta}(r,\theta,t)\hat{\theta} +
E_{\phi}(r,\theta,t)\hat{\phi} \,,
\end{equation}
using spherical coordinates $(r,\theta,\phi)$ with unit vectors
$\hat{r}$, $\hat{\theta}$ and $\hat{\phi}$ and such that $x =
r\cos\phi\sin\theta$, $y = r\sin\phi\sin\theta$, and $z =
r\cos\theta$. The class (\ref{eee}) is the general class of
electric fields compatible with azimuthal symmetry. With an
appropriated scalar potential $V = V(r,\theta,t)$ as well as the
vector potential (\ref{Amono}), it can be proven that
\begin{equation}
F = G\lambda
\end{equation}
is the solution for (\ref{F1}--\ref{F2}). The corresponding
Noether invariant (\ref{basicI}) is
\begin{equation}
\label{I3}
I_3 = \Omega_{3}(y\dot x - x\dot y + g\,z/r) \,,
\end{equation}
gauge independent. This first integral is proportional to the
third component of the Poincar\'e vector (\ref{re}).

If, in addition to the symmetry of rotation about one axis, we
impose the existence of symmetry of rotation about a different
axis, it can be easily proven that the solution for (\ref{Erho0})
is
\begin{equation}
\label{rr}
{\bf E} = E(r,t)\hat{r} \,,
\end{equation}
the general class of central, time-dependent fields. The result is
explained by the $so(3)$ algebra. For instance, the presence of
the extra symmetry of rotations about the $y$ axis imply
rotational symmetry about the $x$ axis, since $[L_{2},L_{3}] = -
L_1$ and by definition the symmetry algebra is closed. In a
similar way to (\ref{I3}), it can be then verified that the Noether
invariants associated to the $SO(3)$ group are the three
components of the Poincar\'e vector. Of course, this is not the
first time that the Poincar\'e vector is shown to be associated to rotational symmetry (see, for instance, reference
\cite{Jackiw}).

\section{Central electric forces}

When the electric field is central, the Poincar\'e vector is
immediately conserved, as a result of the $SO(3)$ symmetry. In
this case, the discussion can be reduced to essentially
one-dimensional, time-dependent motion. To see this, choose axis
so that the conserved Poincar\'e vector may be written ${\bf D} =
(0,0,D)$. Decomposing ${\bf D}$ in components and using spherical
coordinates, we get
\begin{equation}
\cos\theta = - g/D \,, \quad \label{theta}
\dot\phi = D/r^2 \,,
\end{equation}
while the radial component of the Lorentz equation reads, from
(\ref{rr}) and (\ref{theta}),
\begin{equation}
\label{rrr}
\ddot r = E(r,t) + \frac{D^2 - g^2}{r^3} \,.
\end{equation}
Eq. (\ref{theta}) shows that the motion is on a circular
cone whose vertex contains the monopole, and that the angle $\phi$
can be obtained from a simple quadrature once the radial variable
is found from the solution of (\ref{rrr}). This latter equation
involves only $r$ and time.

The presence of extra Noether invariants helps for the integration
of (\ref{rrr}). Since the electric field is central, the category
of Noether symmetries described in subsection III.1 are admitted
if and only if
\begin{equation}
\label{ccc}
{\bf\Omega} = 0 \,, \quad U = U(\bar{r}) \,, \quad \bar{r} = r/\rho \,,
\end{equation}
In the following, we will consider in more detail a case in which
the function $U$ can be conveniently chosen, so that there is an
extra Noether symmetry.

Let us illustrate the initial results of the Section with the electric field
\begin{equation}
\label{Eosc}
{\bf E} = -
\omega^{2}(t){\bf r} + \sigma^{2}{\bf r}/r^4 \,.
\end{equation}
where $\omega(t)$ is an arbitrary function of time and $\sigma$ is
a numerical constant. As observed in the Introduction, for
constant $\omega$ and $\sigma = g$, all bounded trajectories are
periodic for this electric field. In the general, time-dependent
case, (\ref{Eosc}) produces the time-dependent monopole-oscillator
problem, with in addition a repulsive force.

As the electric force is central, $SO(3)$ is known to be admitted
in advance, the Poincar\'e vector being conserved. Besides this
obvious symmetry, there is also symmetry in the form of a
quasi-invariance transformation. To see this, notice that the
scalar potential
\begin{equation}
V = \omega^{2}(t)r^{2}/2 +
\sigma^{2}/2r^2
\end{equation}
can be put in the form (\ref{Vperm}) with $\lambda = 0$ if and only if
\begin{equation}
\label{Unobar} U\left(\frac{r}{\rho}\right) = \frac{1}{2}(\ddot\rho + \omega^{2}(t)\rho)\rho\,r^2 +
\frac{\sigma^{2}\rho^2}{2r^2} \,.
\end{equation}
The right-hand side of (\ref{Unobar}) is properly a function of
$r/\rho$ if and only if $\rho$ satisfies Pinney's \cite{Pinney}
equation
\begin{equation}
\label{pinney}
\ddot\rho + \omega^{2}(t)\rho = k/\rho^3 \,,
\end{equation}
where $k$ is a constant. In this case, we have
\begin{equation}
U(\bar{r}) = \frac{1}{2}k\bar{r}^2 + \frac{\sigma^2}{2\bar{r}^2} \,.
\end{equation}
Noether's invariant (\ref{Iquadra}) is
\begin{equation}
\label{Iq}
I = \frac{1}{2}(\rho\dot{\bf r} - \dot\rho{\bf r})^2 + \frac{k}{2}\left(\frac{r}{\rho}\right)^2 +
\frac{\sigma^2}{2}\left(\frac{\rho}{r}\right)^2 \,.
\end{equation}
In conclusion, the time-dependent monopole oscillator system with
an extra repulsive force do have, besides $SO(3)$ symmetry,
quasi-invariance transformations as Noether symmetries, provided
the function $\rho$ satisfies Pinney's equation.

A more convenient formulation of the invariance properties of the
system is provided by the linearising transform
\begin{equation}
\psi = \rho^2 \,,
\end{equation}
so that Pinney's equation becomes, upon differentiation,
\begin{equation}
\label{linear}
{\buildrel\cdots\over\psi} + 4\omega^{2}\dot\psi + 4\omega\dot\omega\psi = 0 \,.
\end{equation}
The general solution for this last equation is any linear
combination of three independent particular solutions $\psi_1$,
$\psi_2$ and $\psi_3$,
\begin{equation}
\psi = c_{1}\psi_1 + c_{2}\psi_2 + c_{3}\psi_3 \,,
\end{equation}
where $c_1$, $c_2$ and $c_3$ are numerical constants. To each
solution $\psi_i$ correspond one associated Noether point
symmetry, with generator of the form (\ref{ER3}) with $\psi_i =
\rho_{i}^2$,
\begin{equation}
\label{oscgera}
G_i = \psi_{i}\frac{\partial}{\partial t} +
\frac{\dot\psi_{i}{\bf r}}{2}\cdot\frac{\partial}{\partial{\bf r}} \,.
\end{equation}
The associated Noether invariants follows from (\ref{Iq}),
\begin{equation}
\label{Ipsi}
I_i = \frac{1}{2}\left(\psi_{i}\dot{\bf r}^2 - \dot\psi_{i}\,r\dot{r} +
(\frac{\ddot\psi_{i}}{2} + \omega^{2}\psi_{i})\,r^2 +
\frac{\sigma^{2}\psi_{i}}{r^2}\right) \,,
\end{equation}
with $i = 1,2,3$, and where we have eliminated the constant $k$
using Pinney's equation.

The Noether symmetries and invariants can be explicitly shown when
the general solution for (\ref{linear}) is available. In
particular, when
\begin{equation}
\omega = \omega_0 \,,
\end{equation}
a constant, the general solution is
\begin{equation}
\label{psiosc}
\psi = c_1 + c_{2}\cos(2\omega_{0}t) +
c_{3}\sin(2\omega_{0}t) \,.
\end{equation}
The corresponding generators (\ref{oscgera}), obtained for $c_1 =
1$, $c_2 = c_3 = 0$ and cyclic permutations, are
\begin{eqnarray}
G_1 &=& \frac{\partial}{\partial t} \,, \nonumber \\
G_2 &=& \cos(2\omega_{0}t)\frac{\partial}{\partial t} - \omega_{0}\sin(2\omega_{0}t){\bf
r}\cdot\frac{\partial}{\partial{\bf r}} \,,\\
G_3 &=& \sin(2\omega_{0}t)\frac{\partial}{\partial t} + \omega_{0}\cos(2\omega_{0}t){\bf
r}\cdot\frac{\partial}{\partial{\bf r}} \,. \nonumber
\end{eqnarray}
These generators, together with the generators of $SO(3)$,
determine the algebra
\begin{eqnarray}
\left[G_{1},G_{2}\right] &=& - 2\omega_{0}G_3
\quad,\quad \left[G_{2},G_{3}\right] =  2\omega_{0}G_1 \quad,  \nonumber \\
\left[G_{3},G_{1}\right] &=& - 2\omega_{0}G_2 \quad,\quad \left[L_i,L_j\right] =
- \epsilon_{ijk}L_k \quad, \\
\left[G_i,L_j\right] &=& 0 \quad,\quad i,j =
1,2,3 \quad. \nonumber
\end{eqnarray}
Therefore, the Noether point symmetry algebra for the
time-independent monopole-oscillator with an extra repulsive force
have a $so(2,1)\oplus\,so(3)$ structure, the same symmetry algebra
as in the simple magnetic monopole case.

As already seen, $SO(3)$ invariance is associated to the
Poincar\'e vector. On the other hand, invariance under $G_1$,
$G_2$ and $G_3$ corresponds, respectively, to the constants of
motion
\begin{eqnarray}
\label{II1}
I_1 &=& \frac{1}{2}(\dot{\bf r}^2 + \omega_{0}^{2}r^2 + \sigma^{2}/r^2) \,,\\
I_2 &=& \frac{1}{2}\dot{\bf r}^{2}\cos(2\omega_{0}t) + \omega_{0}\,r\dot{r}\sin(2\omega_{0}t) -
\frac{1}{2}\omega_{0}^{2}r^{2}\cos(2\omega_{0}t) +
\frac{\sigma^2}{2r^2}\cos(2\omega_{0}t) \,, \nonumber \\
\label{II3}
I_3 &=& \frac{1}{2}\dot{\bf r}^{2}\sin(2\omega_{0}t) - \omega_{0}\,r\dot{r}\cos(2\omega_{0}t) -
\frac{1}{2}\omega_{0}^{2}r^{2}\sin(2\omega_{0}t) +
\frac{\sigma^2}{2r^2}\sin(2\omega_{0}t) \,. \nonumber
\end{eqnarray}

The six Noether invariants, namely the components of ${\bf D}$ and
$I_i$ above, are not all independent,
since
\begin{equation}
\label{funcional}
\omega_{0}^{2}{\bf D}^2 = I_{1}^2 - I_{2}^2 - I_{3}^2 \,.
\end{equation}

As discussed in the beginning of the Section, the fact that the
electric field is central allows to reduce the problem to the
solution for the radial variable. Here, the existence of the
Noether invariants allows the direct solution for $r(t)$ by
elimination of $\dot{\bf r}$ between the invariants $I_1$, $I_2$ and $I_3$, with the result
\begin{equation}
\label{r}
r^{2}(t) = \frac{1}{\omega_{0}^2}(I_1 - I_{2}\cos(2\omega_{0}t) -
I_{3}\sin(2\omega_{0}t)) \,,
\end{equation}
Inserting $r(t)$ into (\ref{theta}) and integrating, it follows that
the azimuthal variable is
\begin{equation}
\label{p} \phi(t) = \phi_0 +
\arctan\left(\frac{-I_3 + (I_1 + I_2)\tan(2\omega_{0}t)}{\omega_{0}D}\right) \,,
\end{equation}
where $\phi_0$ is a reference angle.

Formulae (\ref{theta}) and (\ref{r}--\ref{p}) are the exact
solution for the time-independent mo\-no\-po\-le-oscillator problem with
an repulsive force. The exact solution involves four independent
integration constants,  $I_1, I_2, I_3$ and $\phi_0$, while $D$ is
functionally dependent on these constants through
(\ref{funcional}). The exact solution does not contain six
integration constants since, from the very beginning, two
components of the Poincar\'e vector were annulled. The remaining
two constants can be easily incorporated, with the price of a less
clear presentation. Finally, it should be stressed that any
time-dependent frequency such that (\ref{linear}) can be exactly
solved leads to exact solution in the same way as the
time-independent case.

\section{Constant magnetic field}

In this Section, we consider the physically relevant case of a
constant magnetic field,
\begin{equation}
\label{b1}
{\bf B} = (0,0,B_{0}) \,,
\end{equation}
where $B_0$ is a numerical constant, and look for Noether point
symmetries and invariants for appropriate electric fields. More
precisely, unlike the magnetic monopole case, we do not search for the
more general class of electric fields for which some Noether point
symmetry is available. In fact, we restrict the treatment to
time-dependent linear electric fields. Such a restriction is again
physically meaningful.

Inserting (\ref{b1}) into (\ref{basicB}) there results an equation
for the vector ${\bf\Omega}$ responsible for time-dependent
rotations,
\begin{equation}
\label{b2}
\dot{\bf\Omega} = \frac{1}{2}{\bf\Omega}\times{\bf B} - \rho\dot\rho{\bf B} \,.
\end{equation}
The above system is easily solved,
\begin{eqnarray}
\Omega_1 &=& c_{1}\cos(B_{0}t/2) + c_{2}\sin(B_{0}t/2) \,,
\nonumber \\ \Omega_2 &=& - c_{1}\sin(B_{0}t/2) +
c_{2}\cos(B_{0}t/2)
\,,  \\
\label{b3} \Omega_3 &=& c_3 - B_{0}\rho^{2}/2 \,, \nonumber 
\end{eqnarray}
where the $c_i$ are integration constants. With the result
(\ref{b3}), the determining equation (\ref{basicE}) for the
electric field is expressed as
\begin{eqnarray}
G\,{\bf E} &=& - 3\rho\dot\rho{\bf E} + {\bf\Omega}\times{\bf E} +
(\rho{\buildrel\cdots\over\rho} + 3\dot\rho\ddot\rho +
B_{0}^{2}\rho\dot\rho){\bf r} + \nonumber \\
\label{b4} &+& ({\bf B}\cdot{\bf
r})(\frac{1}{4}{\bf\Omega}\times{\bf B} - \rho\dot\rho{\bf B}) + \frac{1}{4}({\bf\Omega}\times{\bf
B})\cdot{\bf r}\,{\bf B} + {\bf B}\times\dot{\bf a} + \ddot{\bf a}
\,.
\end{eqnarray}

We have not found the general solution for (\ref{b4}), so that the
general class of admissible electric fields remains to be
determined. However, there is at least one case amenable to exact
calculations, namely the particular case of linear electric fields
of the form
\begin{equation}
\label{b5} {\bf E} = - \omega_{\perp}^{2}(t){\bf r}_{\perp} - \omega_{\parallel}^{2}(t){\bf r}_{\parallel} + {\bf
f}(t) \,,
\end{equation}
where ${\bf r}_{\perp} = (x,y,0)$, ${\bf r}_{\parallel} = (0,0,z)$, ${\bf f} = (f_{1}(t),f_{2}(t),f_{3}(t))$ and
$\omega_{\perp}$, $\omega_{\parallel}$ and the $f_i$ are
time-dependent functions. When the electric field is
linear, both sides of (\ref{b4}) are linear functions of the
coordinates. Equating to zero the coefficients of each coordinate
and of the independent term, the result is a coupled system of
ordinary differential equations for the functions $\rho$ and ${\bf
a}$ composing the symmetry generator,
\begin{eqnarray}
\label{b6}
\rho{\buildrel\cdots\over\rho} + 3\dot\rho\ddot\rho + (B_{0}^{2}
+ 4\omega^{2}_{\perp})\rho\dot\rho
+ 2\omega_{\perp}\dot\omega_{\perp}\rho^2 &=& 0 \,, \\
\label{b7}
\rho{\buildrel\cdots\over\rho} + 3\dot\rho\ddot\rho +
4\omega^{2}_{\parallel}\rho\dot\rho +
2\omega_{\parallel}\dot\omega_{\parallel}\rho^2 &=& 0 \,, \\
\label{b8}
\ddot{a_1} + \omega_{\perp}^{2}\,a_1 &=& B_{0}\dot{a}_2 + d_{1}(t) \,, \\
\label{b9}
\ddot{a_2} + \omega_{\perp}^{2}\,a_2 &=& - B_{0}\dot{a}_1 + d_{2}(t) \,, \\
\label{b10}
\ddot{a_3} + \omega_{\parallel}^{2}\,a_3 &=& d_{3}(t) \,.
\end{eqnarray}
Moreover, the following algebraic relations must be satisfied,
\begin{equation}
\label{b11}
\Omega_{1}(\omega_{\parallel}^2
- \omega_{\perp}^2 - B_{0}^{2}/4)
= \Omega_{2}(\omega_{\parallel}^2 - \omega_{\perp}^2 - B_{0}^{2}/4) = 0 \,.
\end{equation}
In (\ref{b8}-\ref{b10}), the vector ${\bf d} = (d_{1}, d_{2},
d_{3})$ is defined according to
\begin{equation}
\label{b12}
{\bf d} = \rho^{2}\dot{\bf f} + 3\rho\dot\rho{\bf f} + {\bf f}\times{\bf\Omega} \,.
\end{equation}

Once the system (\ref{b6}-\ref{b11}) is solved, the associated
Noether invariant is found from (\ref{basicI}), which require both
the electromagnetic potentials and the function $F$, solution for
(\ref{F1}-\ref{F2}). The electromagnetic potentials are
\begin{eqnarray}
\label{b13}
{\bf A} &=& \frac{B_0}{2}(-y,x,0) + \nabla\lambda({\bf r},t) \,, \\
\label{b14}
V &=& \frac{1}{2}\omega^{2}_{\perp}(t)(x^2 + y^2) +
\frac{1}{2}\omega^{2}_{\parallel}(t)\,z^2
- {\bf f}(t)\cdot{\bf r} - \frac{\partial\lambda}{\partial t}({\bf r},t) \,,
\end{eqnarray}
where $\lambda({\bf r},t)$ is an arbitrary gauge function. The
function $F$, the last ingredient for the Noether invariant,
follows from the use of these electromagnetic potentials in the
system (\ref{F1}-\ref{F2}), and reads
\begin{equation}
\label{b15}
F = \frac{1}{2}(\rho\ddot\rho
+ {\dot\rho}^2)r^2 + \dot{a}\cdot{\bf r}
+ \frac{1}{2}{\bf B}\cdot{\bf a}\times{\bf r}
+ \int^{t}d\mu\,{\bf f}(\mu)\cdot{\bf a}(\mu) + G\,\lambda \,.
\end{equation}
The Noether invariant (\ref{basicI}) is then expressed as
\begin{eqnarray}
I &=& \frac{1}{2}(\rho\dot{\bf r} - \dot\rho{\bf r})^2
+ {\bf\Omega}\cdot\dot{\bf r}\times{\bf r}
+ \frac{\rho^2}{2}(\omega_{\perp}^{2}(x^2 + y^2)
+ \omega_{\parallel}^{2}z^{2}) + \frac{1}{2}\rho\ddot\rho\,r^2 + \nonumber \\
&+&\frac{1}{2}{\bf\Omega}\cdot({\bf B}\times{\bf r})\times{\bf r}
-{\bf a}\cdot\dot{\bf r} +\dot{\bf a}\cdot{\bf r}
-{\bf a}\cdot{\bf B}\times{\bf r} + \nonumber \\
&+& \label{b16} \int^{t}d\mu{\bf
f}(\mu)\cdot{\bf a}(\mu) \,,
\end{eqnarray}
gauge independent as it must be. For the explicit form of the
Noether invariant, we have to solve the system
(\ref{b6}-\ref{b11}) giving the functions $\rho$ and $a_i$ which
remain to be obtained.

After a detailed but straightforward analysis, we distinguish two
classes of solutions for the system (\ref{b6}-\ref{b11}),
according to the functions $\omega_{\perp}$ and
$\omega_{\parallel}$ entering the electric field and the magnetic
field strength. The two classes of solutions are treated
separately.

\subsection{The $\omega_{\parallel}^{2}(t) = \omega_{\perp}^{2}(t) + B_{0}^{2}/4 $ case}

In the situation where the relation $\omega_{\parallel}^{2}(t) =
\omega_{\perp}^{2}(t) + B_{0}^{2}/4$ is valid, the condition
(\ref{b11}) becomes an identity, so that the components $\Omega_1$
and $\Omega_2$ are left free. Referring to equation (\ref{b3})
defining the vector ${\bf\Omega}$ for a constant magnetic field,
this means that the constants $c_1$ and $c_2$ are left free.
Moreover, the equations (\ref{b6}) and (\ref{b7}) are identical, becoming, using the linearising transform $\psi = \rho^2$,
\begin{equation}
\label{b17} {\buildrel\cdots\over\psi} +
4\omega_{\parallel}^{2}\dot\psi +
4\omega_{\parallel}\dot\omega_{\parallel}\psi = 0 \,,
\end{equation}
a third-order linear equation. Denote the general solution as
\begin{equation}
\psi = c_{4}\psi_1 + c_{5}\psi_2 + c_{6}\psi_3 \,,
\end{equation}
where $c_i$ are numerical constants and $\psi_i$ independent
particular solutions. Taking into account the three arbitrary
numerical constants entering ${\bf\Omega}$, plus $c_4$, $c_5$,
$c_6$ and the six integration constants for the system
(\ref{b8}-\ref{b10}), we arrive at a 12-parameter group of Noether
point symmetries.

The construction of the symmetry group is best explained with a
concrete example. In the last part of the Section, let us study in
more detail the case
\begin{equation}
\label{b18} {\bf f} = 0 \,, \quad \dot\omega_{\perp} =
\dot\omega_{\parallel} = 0 \,,
\end{equation}
that is, the cases of time-independent harmonic fields with no forcing term. In
this context, (\ref{b17}) is easily solved, giving
\begin{equation}
\label{b19}
\psi = c_4 + c_{5}\cos(2\omega_{\parallel}t) + c_{6}\sin(2\omega_{\parallel}t) \,.
\end{equation}
In addition, the system (\ref{b8}-\ref{b10}) has the general solution
\begin{eqnarray}
\label{b20}
a_1 &=& c_{7}\cos(\omega_{1}t) + c_{8}\sin(\omega_{1}t)
+ c_{9}\cos(\omega_{2}t) - c_{10}\sin(\omega_{2}t) \,, \nonumber \\
\label{b21}
a_2 &=& - c_{7}\sin(\omega_{1}t) + c_{8}\cos(\omega_{1}t)
+ c_{9}\sin(\omega_{2}t) + c_{10}\cos(\omega_{2}t) \,,\\
\label{b22}
a_3 &=&  c_{11}\cos(\omega_{\parallel}t)
+ c_{12}\sin(\omega_{\parallel}t) \,, \nonumber
\end{eqnarray}
where $c_i$ are integration constants and
\begin{equation}
\label{b23} \omega_1 = \frac{1}{2}(B_0 + \sqrt{B_{0}^2 +
4\omega_{\perp}^2}) \,, \quad \omega_2 = \frac{1}{2}(- B_0 +
\sqrt{B_{0}^2 + 4\omega_{\perp}^2}) \,.
\end{equation}

Choosing $c_i = \delta_{ij}$, for $i = 1,...,12$ and $j =
1,...,12$, the twelve Noether point symmetry generators can be
constructed from (\ref{geral}), (\ref{b3}) and
(\ref{b19}-\ref{b22}). They read
\begin{eqnarray}
G_1 &=& \cos(B_{0}t/2)(y\partial/\partial z - z\partial/\partial y)
- \sin(B_{0}t/2)(z\partial/\partial x - x\partial/\partial z) \,,  \nonumber \\
G_2 &=& \sin(B_{0}t/2)(y\partial/\partial z - z\partial/\partial y) +
 \cos(B_{0}t/2)(z\partial/\partial x - x\partial/\partial z) \,, \nonumber \\
G_3 &=& x\partial/\partial y - y\partial/\partial x \,, \quad G_4 = \partial/\partial t
- (B_{0}/2)(x\partial/\partial y - y\partial/\partial x) \,, \nonumber \\
G_5 &=& \cos(2\omega_{\parallel}t)\,(\partial/\partial t
- (B_{0}/2)(x\partial/\partial y - y\partial/\partial x))
- \omega_{\parallel}\sin(2\omega_{\parallel}t)\,{\bf r}\cdot\nabla \,, \nonumber \\
G_6 &=& \sin(2\omega_{\parallel}t)\,(\partial/\partial t - (B_{0}/2)(x\partial/\partial y
- y\partial/\partial x)) + \omega_{\parallel}\cos(2\omega_{\parallel}t)\,{\bf r}\cdot\nabla \,, \nonumber \\
G_7 &=& \cos(\omega_{1}t)\partial/\partial x - \sin(\omega_{1}t)\partial/\partial y \,, \nonumber \\
\quad G_8 &=& \sin(\omega_{1}t)\partial/\partial x + \cos(\omega_{1}t)\partial/\partial y \,, \nonumber \\
G_9 &=& \cos(\omega_{2}t)\partial/\partial x + \sin(\omega_{2}t)\partial/\partial y \,,
\\ G_{10} &=&
- \sin(\omega_{2}t)\partial/\partial x + \cos(\omega_{2}t)\partial/\partial y \,, \nonumber \\
G_{11} &=& \cos(\omega_{\parallel}t)\partial/\partial z \,, \quad
G_{12} = \sin(\omega_{\parallel}t)\partial/\partial z \,. \nonumber
\end{eqnarray}

The conserved quantities associated to the above generators
follows from (\ref{b16}),
\begin{eqnarray}
I_1 &=& \cos(B_{0}t/2)(z(\dot{y} + B_{0}x/2) - y\dot{z}) + \sin(B_{0}t/2)(z(\dot{x}
- B_{0}y/2) - x\dot{z}) \,, \nonumber \\
I_2 &=& \sin(B_{0}t/2)(z(\dot{y} + B_{0}x/2) - y\dot{z}) - \cos(B_{0}t/2)(z(\dot{x}
- B_{0}y/2) - x\dot{z}) \,, \nonumber \\
I_3 &=& x\dot{y} - y\dot{x} + (B_{0}/2)(x^2 + y^2) \,, \nonumber \\
I_4 &=& {\dot{\bf r}}^{2}/2 + (B_{0}/2)(x\dot{y} - y\dot{x}) + (1/2)((\omega_{\perp}^2
+ B_{0}^{2}/2)(x^2 + y^2) + \omega_{\parallel}^{2}z^{2}) \,, \nonumber \\
I_5 &=& (1/2){\dot{\bf r}}^{2}\cos(2\omega_{\parallel}t)
+ \omega_{\parallel}r\dot{r}\sin(2\omega_{\parallel}t) + \nonumber \\
&+& (B_{0}/2)(x\dot{y} - y\dot{x})\cos(2\omega_{\parallel}t) \nonumber
- (1/2)(\omega_{\perp}^{2}(x^2 + y^2) + \omega_{\parallel}^{2}z^{2})\cos(2\omega_{\parallel}t)
\,, \\
I_6 &=& (1/2){\dot{\bf r}}^{2}\sin(2\omega_{\parallel}t)
- \omega_{\parallel}r\dot{r}\cos(2\omega_{\parallel}t) + \nonumber \\
&+& (B_{0}/2)(x\dot{y} - y\dot{x})\sin(2\omega_{\parallel}t)
- (1/2)(\omega_{\perp}^{2}(x^2 + y^2)
+ \omega_{\parallel}^{2}z^{2})\sin(2\omega_{\parallel}t)  \,, \nonumber \\
I_7 &=& (\dot{x} + \omega_{2}y)\cos(\omega_{1}t) - (\dot{y}
- \omega_{2}x)\sin(\omega_{1}t) \,, \nonumber \\
I_8 &=& (\dot{x} + \omega_{2}y)\sin(\omega_{1}t) + (\dot{y}
- \omega_{2}x)\cos(\omega_{1}t) \,, \\
I_9 &=& (\dot{x} - \omega_{1}y)\cos(\omega_{2}t) + (\dot{y}
+ \omega_{1}x)\sin(\omega_{2}t) \,, \nonumber \\
I_{10} &=& - (\dot{x} - \omega_{1}y)\sin(\omega_{2}t) + (\dot{y}
+ \omega_{1}x)\cos(\omega_{2}t) \,, \nonumber \\
I_{11} &=& \dot{z}\cos(\omega_{\parallel}t)
+ \omega_{\parallel}z\sin(\omega_{\parallel}t) \,, \quad I_{12}
= \dot{z}\sin(\omega_{\parallel}t) - \omega_{\parallel}z\cos(\omega_{\parallel}t) \,. \nonumber
\end{eqnarray}

The invariants $I_i$ for $i = 7,...12$ are sufficient, from elimination of the velocities, for the
general solution of the equations of motion,
\begin{eqnarray}
x &=& \frac{1}{2\omega_{\parallel}}(I_{7}\sin(\omega_{1}t) - I_{8}\cos(\omega_{1}t)
+ I_{9}\sin(\omega_{2}t) + I_{10}\cos(\omega_{2}t)) \,, \nonumber \\
y &=& \frac{1}{2\omega_{\parallel}}(I_{7}\cos(\omega_{1}t) + I_{8}\sin(\omega_{1}t)
- I_{9}\cos(\omega_{2}t) + I_{10}\sin(\omega_{2}t)) \,, \\
z &=& \frac{1}{\omega_\parallel}(I_{11}\sin(\omega_{\parallel}t)
- I_{12}\cos(\omega_{\parallel}t)) \,. \nonumber
\end{eqnarray}
The remaining invariants $I_i$ for $i = 1,...6$ are functionally
dependent on the invariants associated to time-dependent
translations.

\subsection{The $\omega_{\parallel}^{2}(t) \neq
\omega_{\perp}^{2}(t) + B_{0}^{2}/4 $ case}

In this situation, condition (\ref{b11}) is satisfied only if
$\Omega_1 = \Omega_2 = 0$. This rules out rotational symmetries
about the $x$ and $y$ axis, and imply $c_1 = c_2 = 0$ in (\ref{b3}). In
addition, equations (\ref{b6}) and (\ref{b7}) are equivalent only
if $\rho = c_4$, a numerical constant. In fact, it is possible, in
principle, to have non constant solutions for $\rho$ satisfying
both (\ref{b6}-\ref{b7}). However, this possibility is allowed
only if $\omega_\parallel$ and $\omega_\perp$ are related by a
somewhat complicated relation which we refrain from writing here.
In conclusion, when $\omega_{\parallel}^{2}(t) \neq
\omega_{\perp}^{2}(t) + B_{0}^{2}/4$ generically there is a
8-parameter group of Noether point symmetries. This group
comprises rotations about the $z$ axis, time-translation and the
six time-dependent space translations determined by the solution
of (\ref{b8}-\ref{b10}). Accordingly, in the time-independent case
with no forcing term, specified by (\ref{b18}), the symmetries
generated by $G_2$, $G_3$, $G_5$ and $G_6$ are lost, as well as
the associated Noether invariants. Even if the symmetry structure
is less rich, it is sufficient for the general solution of the
equations of motion, since the six time-dependent translational
symmetries are not broken.

\section{Conclusion}

We have obtained the system of partial differential equations to
be satisfied by the electromagnetic field so that the action
functional for non-relativistic motion is invariant under
continuous point transformations. This system of equations has
been completely solved when the magnetic field is produced by a fixed
magnetic monopole. The associated constants of motion can have
linear or quadratic dependencies on the velocity. These constants
of motion can be used to integrate the Lorentz equations, as in
the monopole-oscillator problem with an extra repulsive force
field. Moreover, we have treated the case of a constant magnetic
field plus a harmonic electric field with a forcing term.

The main open problem that still deserves attention is the
complete solution of the basic system (\ref{basicB}-\ref{basicE})
for the electromagnetic field. The technical drawback here is the
determination of the canonical group coordinates for the generator $G$
in (\ref{geral}) with arbitrary $\rho(t)$, ${\bf\Omega}(t)$ and
${\bf a}(t)$. It can be verified that this issue can be solved at
least when ${\bf\Omega}(t)$ has a fixed direction. Other
particular solutions may be valuable. In addition, the system
(\ref{basicB}-\ref{basicE}) is worth to be considered in the case
of other particular electromagnetic fields, for which the magnetic
field is not in the form of a magnetic monopole or of a constant
field.

Other direction of research is the search for Lie point symmetries
for non-relativistic charged particle motion under generic
electromagnetic fields. In the two-dimensional case, this problem
can be completely solved \cite{H2}. In three dimensions, certainly
the difficulties are greater than those we are faced in the case
of Noether point symmetries, since the Noether group of point
symmetries is a subgroup of the Lie group of point symmetries.
Again, the stumbling block is the finding of canonical group
coordinates for the generator of Lie symmetries. Finally, further
extensions involve the use of transformations of more general
character, such as dynamical or nonlocal transformations.

\medskip

\noindent{{\bf Acknowledgements}}

This work has been supported by the Brazilian agency
Conselho Nacional de Desenvolvimento Cient\'{\i}fico e Tecnol\'ogico (CNPq).

\end{document}